\documentclass[aps,caption,twocolumn,amsmath,amssymb,superscriptaddress,showpacs]{revtex4-1}
\usepackage{graphicx}
\usepackage{color,soul}
\usepackage{afterpage}
\usepackage[final]{pdfpages}

\makeatletter
\AtBeginDocument{\let\LS@rot\@undefined}
\makeatother

\begin{document}
\title{Local Pore Size Correlations Determine Flow Distributions in Porous Media}
\author{Karen Alim}
\email{karen.alim@ds.mpg.de}
\affiliation{John A.~Paulson School of Engineering and Applied Sciences and Kavli Institute for Bionano Science and Technology, Harvard University, Cambridge, MA 02138, USA}
\affiliation{Max Planck Institute for Dynamics and Self-Organization, 37077 G\"ottingen, Germany}
\author{Shima Parsa}
\affiliation{John A.~Paulson School of Engineering and Applied Sciences and Kavli Institute for Bionano Science and Technology, Harvard University, Cambridge, MA 02138, USA}
\author{David A. Weitz}
\affiliation{John A.~Paulson School of Engineering and Applied Sciences and Kavli Institute for Bionano Science and Technology, Harvard University, Cambridge, MA 02138, USA}
\affiliation{Department of Physics, Harvard University, Cambridge, MA 02138, USA}
\author{Michael P. Brenner}
\affiliation{John A.~Paulson School of Engineering and Applied Sciences and Kavli Institute for Bionano Science and Technology, Harvard University, Cambridge, MA 02138, USA}
\begin{abstract}
The relationship between the microstructure of a porous medium and the observed flow distribution is still a puzzle. We resolve it with an analytical model, where the local correlations between adjacent pores, which determine the distribution of flows propagated from one pore downstream, predict the flow distribution. Numerical simulations of a two-dimensional porous medium verify the model and clearly show the transition of flow distributions from  $\delta$-function-like via Gaussians to exponential with increasing disorder. Comparison to experimental data further verifies our numerical approach.
\end{abstract}
\maketitle
 Fluid flow through a porous medium is an important problem for many technological applications ranging from oil recovery to chemical reactors \cite{Bear:88,Dullien:79,Sahimi:95,Wong:1988}. Flows are characterized by heterogeneous flow patterns, arising due to complexity on pore network scale \cite{Koch1988,Seymour2004,Berkowitz2006} as well as pore scale \cite{Bijeljic2011,Edery2016}. Large scale flow dynamics such as permeability and flow dispersion are well modeled based on medium-wide measures like pore size distribution and local conductance \cite{Fatt1956,Katz1986,Sahimi1993,Whitaker1967}.
Thus, flow characteristics on scales much larger than a pore size are well understood, however, there has been significant controversy about the flow distributions on the scale of individual pores \cite{Northrup:93,Cenedese:96, Rashidi:96,Lebon:96,Kutsovsky:96,Shattuck:97,Maier:98,Maier:99,Sederman:01,Moroni:01,Magnico:03,Talon:03,Araujo:06,Lachhab:08,Huang:08,Hassan:08,Sen:12,Datta:13}.   
The observed flow velocity distributions range from Gaussian \cite{Maier:98,Araujo:06}, to lognormal \cite{Cenedese:96,Talon:03,Lachhab:08} to exponential \cite{Lebon:96,Kutsovsky:96,Shattuck:97,Maier:98,Maier:99,Sederman:01,Moroni:01,Datta:13,Matyka:2016}.   There is little understanding of why these different distributions are observed, or what causes a particular distribution to occur in a given situation. Current models predict the flow velocity distribution from the pore size distribution \cite{Kutsovsky:96,Maier:99}, describing the porous medium by a bundle of parallel capillaries with the given pore sizes \cite{Scheidegger:57}. The flow velocity distribution then follows directly from the pore size distribution. While this simple mapping predicts an exponential flow velocity distribution for a Gaussian pore size distribution, it often fails to account for other distributions. For example it has been observed that two different porous media with very similar Gaussian-like pore size distributions have very different flow velocity distributions, namely, Gaussian and exponential \cite{Maier:98}.

\begin{figure}[htbp]
\centering{\includegraphics[width=0.5\textwidth]{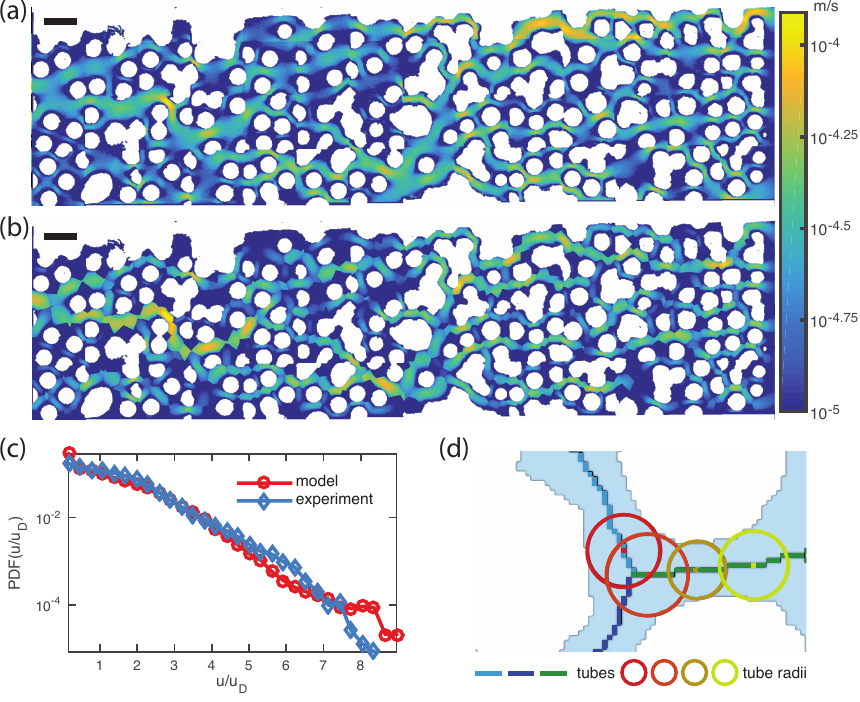}}
\caption{Example of exponential distribution of flow velocities faithfully modeled as flow through a network of tubes with varying radii. (a) Map of measured  absolute flow velocities in two-dimensional pore space (scale bar $500 \mu m$).  Solid phase overlayed in white is a reprint of a slice through a three-dimensional pore space. (b) Numerically predicted Poiseuille flow profile based on a network of tubes of varying radius. (c) Exponential distribution of flow velocities $u$ normalized by the Darcy flow velocity $u_D$ of maps shown in (a) and (b). Note that there is no adjustable parameter. (d) Illustration of how pore space is mapped to a network of tubes (cool colors) with varying tube radius (hot colors).}
\label{fig_exp}
\end{figure}

In this Letter we present a model that explains why different flow distributions occur in different situations. The essential ingredient of our model is not the pore size distribution itself but instead local correlations between adjacent pores; these determine the distribution of fractions of fluid flow propagated from one pore to others that are downstream. We show that when there is sufficient disorder at the pore scale, these flow rates are partitioned randomly. Analytical arguments based on the so-called q model, originally proposed to describe force propagation in bead packs \cite{Liu:95, Coppersmith:96}, then imply that the flow distribution is exponential. We verify these conclusions using numerical simulations of low Reynolds number fluid flow through a two-dimensional model porous medium, formed of circular or elliptical discs placed on a square or triangular lattice with increasing disorder.  The simulations quantitatively reproduce flow distributions from experiments of a two-dimensional porous medium, and clearly show a transition of  two $\delta$ like distributions to Gaussians to an exponential distribution of flows with increasing disorder in the ordering of the discs. We find that pore size distributions are a bad predictor of flow distributions in contrast to the distributions of fraction of propagated fluid flow.

We begin by presenting numerical simulations of fluid flow in a two-dimensional porous medium.
To do this, we reduce the full pore space to a network of connected tubes, where each tube's diameter narrows and widens according to the existing pore space. The flow within each tube of rectangular cross section is calculated by solving for Kirchoff's circuit law for two-dimensional Poiseuille flow within the tubes of varying diameter. This model is an approximation of the full two-dimensional Stokes flow representing the low Reynolds number flow within the porous medium. To verify that flow through a porous medium can be modeled as flow through a network of tubes we compare the predictions from the model to
experimental measurements of flow velocities in a two-dimensional porous medium.
For the experimental data of a  two-dimensional micromodel of porous media we chose a pattern obtained from the optical slices in a three-dimensional, loose packing of glass beads with a confocal microscope \cite{Krummel2013} as well as numerically generated patterns. The two-dimensional micromodel is an array of pillars with an average diameter of 300 $\mu$m and height of 120 $\mu$m. We use standard soft lithography technique \cite{Xia1998} to fabricate the micromodel in PDMS, poly(dimethylsiloxane).  The fluid, a mixture of 84 w\% Glycerol and water is seeded with 1 $\mu$m latex fluorescent tracer particles at 0.006 vol\%. We use particle image velocimetry \cite{Cenedese:96, Datta:13} for measuring the flow velocities. The magnitude of the velocities is measured at a volumetric flow rate of 9 $\mu$L/hr.
Using these parameters, and the geometry of the porous medium, we numerically predict the resulting distribution of flow velocities. The overall pattern of predicted flow velocities is very similar to the experimentally observed, as shown by the two panels in Fig.~\ref{fig_exp}. Slight deviations arise, since the underlying porous medium pattern used in the model has artifacts arising from the stitching of microscope images. Also very small velocities are not well resolved since the fluorescent tracers do not enter regions of vanishingly small flow.  Despite these deviations the histogram of measured flows quantitatively compares well with those predicted, see Fig.~\ref{fig_exp}. Note, that there is no adjustable parameter.
\begin{figure}[htbp]
\centering{\includegraphics[width=0.45\textwidth]{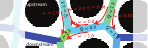}}
\caption{Example of normalized flows $Q/\mathrm{mean}(Q)$ (black) and fractions of propagated flows $\omega$ (red). At a node point flows from incoming tubes slit up in fractions, for example $\omega=\{1.0,0.0\}$ or $\omega=\{0.47,0.52\}$, top left and right, respectively. Kirchhoff's law ensures that fractions sum up to one. Flow in a tube downstream sums up flow contributions from tubes upstream, for example $2.2=1 \times 1.3 +1 \times 0.9$, bottom left.}
\label{fig_weightcartoon}
\end{figure}

\begin{figure*}[htbp]
\centering{\includegraphics[width=1\textwidth]{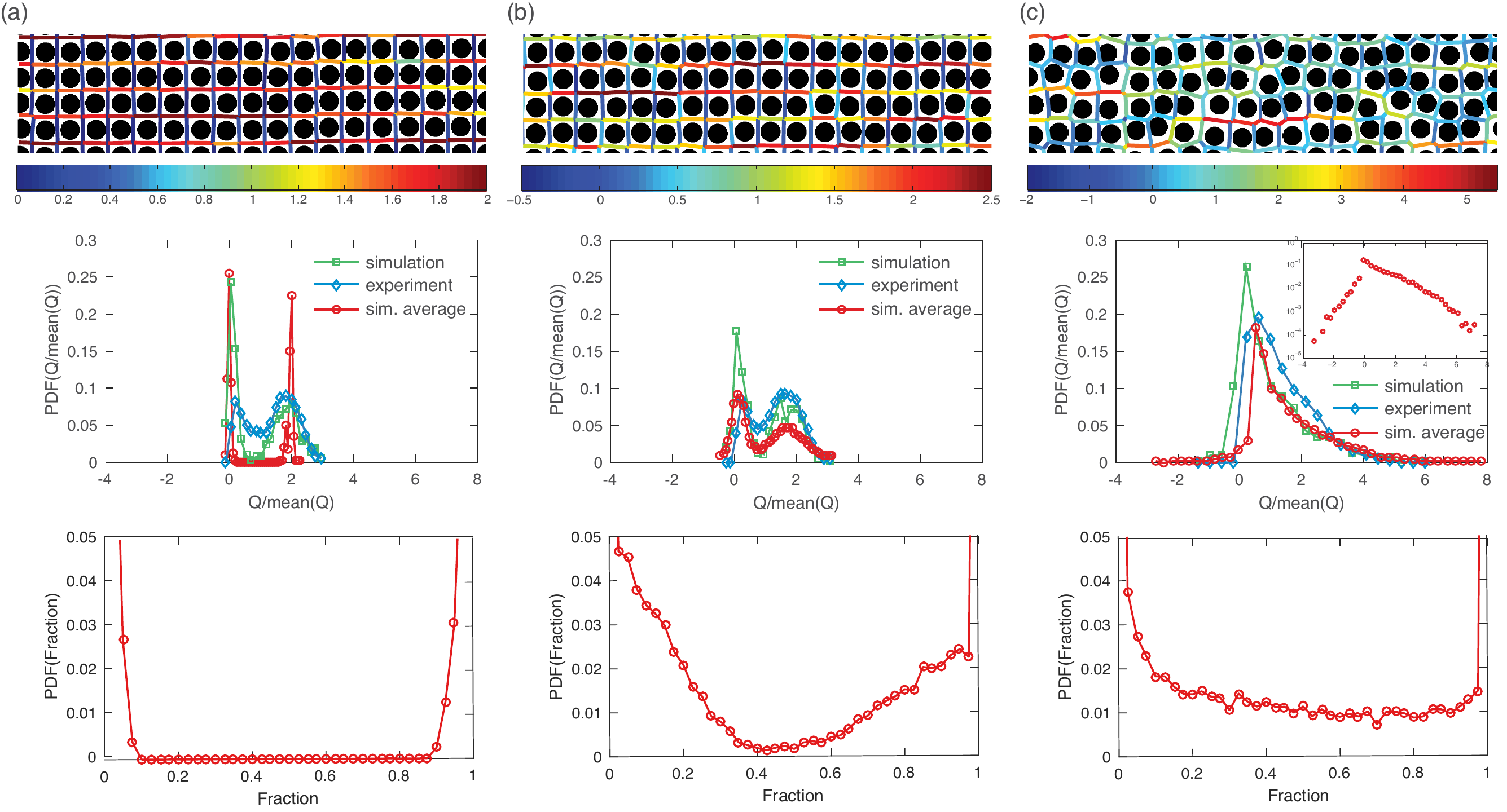}}
\caption{Flow statistics for increasing packing disorder. Snap shot of flow pattern ({\it first row}, note individual scale bars), flow distribution ({\it second row}) and the distribution of the fraction of propagated flows ({\it third row}) for three different packings of increasing disorder $\sigma = 0.01$ (a), $\sigma = 0.1$ (b), and $\sigma = 0.5$  (c).  With the increase in disorder the two delta-like flow distributions in the regular packing, spread out into Gaussians of unequal weight and eventually built up an exponential distribution. This transition is accompanied with the fraction of propagated flow rates spreading out from zero and one and eventually homogeneously populating fractions between the two limits. Comparison of experimentally measured flow rates ({\it blue}) of the micromodels in (a), corresponding numerical solution ({\it green}), and averages of 20 independent numerical runs ({\it red}). Note, that the optical measurement does not allow for good resolution of small velocities. Also, the production and imaging process adds disorder broadening the flow distribution particularly at $\sigma = 0.01$, accounted for by the simulation results based on the experimental micromodel pattern.}
\label{fig_flowstats}
\end{figure*}

 Now that we established the reduction of a porous medium to a network of tubes, we turn to explore numerically generated porous media of varying disorder. Discs of diameter $D=120 \mathrm{px}$, $\mathrm{px}$ short for pixels, are placed on the grid points $(x,y)$ of a square grid allowing for some disorder of magnitude $\sigma D$ along both $x$ and $y$ direction. We choose the porosity in the rectangular space of $1200 \mathrm{px} \times 4800 \mathrm{px}$ to be $\phi=0.5$. We expect the network of tubes model to break down at very high porosities with pore spaces much larger than obstacle size but to hold at least for porosities up to the porosity $\phi=0.6$ of our benchmark comparison shown in Fig.~\ref{fig_exp}.  In simulations the resulting pore space is skeletonized into a network of tubes of varying diameter, with periodic boundary conditions along the short axis. Tube diameters at each point along the tube's skeleton are determined as the minimal distance of each point to the porous media. Each tube is oriented toward its node closest to the outflow along $x$.  
 Flow orientation is determined by being a parallel, positive sign, or anti-parallel, negative sign, to a tube's orientation. Because of this definition negative flow rates arise that are a mere artifact of the orientation of tubes in a Cartesian coordinate system but do not indicate back flows. The flow rates $Q$ at every pixel along the pore space network are tabulated to give the probability distribution function (PDF) of the flow rates for a given pore space geometry, see Fig.~\ref{fig_flowstats}.   Similarly the fraction of flow $\omega$ propagated from one tube onward to its neighboring tubes at every network node is measured.

At low disorder $\sigma=0.01$, when the pore space forms a square grid the flow distribution displays two $\delta$-distributions, one at zero for all pore tubes perpendicular to the pressure drop, and one at the normalized flow $Q/\mathrm{mean}(Q)=2$ corresponding to the uniform flow in the pores parallel to the pressure drop. Upon increasing the disorder to $\sigma=0.1$ the two $\delta$-functions broaden out, with the one previously centered at zero remaining more peaked than the other. Already at a disorder amplitude of half a radius, $\sigma=0.5$, the probability distribution of the flow rates is exponential and remains so for higher disorder. In our simulations, the observed stretched exponential distribution of flux rates as defined in Ref.~\cite{Araujo:06} is equivalent to an exponential distribution of flow rates. We observe the same transition from $\delta$ like distribution via Gaussian to exponential distribution with increasing disorder in experimentally measured flow rates through porous media with patterns drawn from our numerical simulation. The transition is also robust against variation in obstacle shape or underlying grid topology, as long as no additional disorder is introduced that would expedite the transition, see Supplemental Material \cite{supplement}. Why should we expect an exponential distribution of flow rates and hence velocities? What characteristic of the pore space determines the flow distribution? To answer these questions we turn to an analytical description of the local flow dynamics.

Conservation of mass, in fluid dynamics stated as Kirchhoff's circuit law, governs the local dynamics of flows at a node point within the pore space network: the amount of flow into a node has to be equal to the amount of flow flowing out of the node. Hence, the flow rate $Q_j$ in a tube $j$ is the sum of all contributing flows from all $N$ tubes connected to $j$ upstream, each contributing a fraction $\omega_{ij}$ of their total flow $Q_i$ to that tube, see Fig.~\ref{fig_weightcartoon},
\begin{equation}
Q_j = \sum_{i=1}^N \omega_{ij} Q_i.
\label{eqn_stochastic}
\end{equation}
Kirchhoff's law sets the sum over all fractions propagated to obey $\sum_{i}\omega_{ij}=1$. Any set of flow fractions denoted $\eta(\omega)$ is normalized $\int_0^1d\omega\,\eta(\omega) = 1$ and satisfies $\int_0^1d\omega\,\omega\,\eta(\omega) = 1/N$, since we consider a homogeneous connectivity with $N$ tubes connecting at a node point. This normalization is applicable if noncontributing flows are accounted for with a zero weight, see Fig.~\ref{fig_weightcartoon}. This description of local flow propagation is very similar to how a layer of beads supports the weight of the beads above them in random bead packs \cite{Liu:95, Coppersmith:96}.
To solve for the distribution of flow rates $P(Q)$ given a distribution of fractions $\eta(\omega)$ a mean field description analogous to the bead pack model is formulated, see similar calculation in Ref.~cite{Coppersmith:96}. The spatial velocity-velocity correlation has been shown to decay exponentially with the distance. The correlation decreases significantly with decreasing porosity \cite{Araujo:06}, supporting a mean field approximation to be valid for low porosities.
Thus, ignoring correlations of flow rates in tubes upstream of tube $j$, Eq.~\eqref{eqn_stochastic} gives rise to a recursive relation for the flow distribution function $P(Q)$,
\begin{eqnarray}
P(Q_j) = &&\prod_{i=1}^{N}\left\{\int_0^1d\omega_{ij}\;\eta(\omega_{ij})\int_0^{\infty}dQ_i\;P(Q_i)\right\}\nonumber\\
&&\delta\left(\sum_{i=1}^N\omega_{ij}Q_i-Q_j\right).
\label{eqn_meanfield}
\end{eqnarray}
We normalize the flows in the pore space by their mean, defining $\int_0^{\infty}dQ\,Q\,P(Q)=1$.

To understand the consequences of these formulae, it is illustrative to explore the case of flows in a pore space forming a square grid. Here, tubes along the pressure gradient propagate their flow onward with a fraction of $1$ to the next tube along the pressure gradient giving rise to a $\delta$ distribution at just that flow rate. Tubes perpendicular to the flow do not receive any flow $\omega_{ij}=0$ from their neighbors constituting a second $\delta$ function in $P(q)$ at $q=0$. The distribution of the fractions $\eta(\omega)$ is  a sum of two $\delta$ functions  at $\omega_{ij}=0,1$. The key insight is that the distribution of fractions fully defines the flow distribution.

When we turn to the case of highest disorder considered in the simulation, $\sigma = 0.5$, see Fig.~\ref{fig_flowstats}(c), we observe that in addition to two $\delta$-function-like contributions at $\omega_{ij}=0,1$ all fractions between these two limits are equally populated at a roughly constant rate $\alpha$, i.e.,
\begin{equation}
\eta(\omega) = \alpha +\frac{1-\alpha}{2} \delta(\omega) +\frac{1-\alpha}{2}\delta(\omega-1),
\end{equation}
where the prefactors of the $\delta$ functions ensure that the normalization and average mean value are satisfied. The distribution of fractions is the superposition of two flow patterns at the three-tube nodes. Either, there are two inflows and one outflow, in which case flow fractions are either $0$ or $1$ resulting in the two $\delta$-functions, or there is one inflow that splits up into two outflow. The fraction, in which the inflow splits up is at high disorder random and as such gives rise to a uniform distribution of fractions.

 The aim of the following deduction is to show that such a distribution of fraction gives rise to an exponential flow distribution. To this end we first take the Laplace transform of the constituting equation for the flow distribution Eq.~\eqref{eqn_meanfield}, resulting in
\begin{equation}
\tilde{P}(s)= \left[\int_0^1d\omega\;\eta(\omega)\tilde{P}(\omega s)\right]^N.
\label{eqn_laplace}
\end{equation}
In the case of the disordered packing the topology of the pore space network forms to over 99\% nodes with two incoming and one outgoing flows or vice versa, see \cite{supplement}. We therefore restrict our calculation to the case of $N=2$. To simplify Eq.~\eqref{eqn_laplace} we define $\tilde{V}(s)=[\tilde{P}(s)]^{1/2}$, multiply the result by $s$ and differentiate by $s$, yielding,
\begin{eqnarray}
\lefteqn{\frac{1+\alpha}{2}\tilde{V}^2(s)-\tilde{V}(s)+\frac{1-\alpha}{2}\tilde{V}(0)=}\nonumber\\
&&\left[1-(1-\alpha)\tilde{V}(s)\right]s\frac{d\tilde{V}(s)}{ds}.
\label{eqn_laplaceV}
\end{eqnarray}
We find that $\tilde{V}(s)$ is a quickly decaying function, approaching $\tilde{V}(s\to\infty)=\frac{1-\alpha}{1+\alpha}$. We can analytically solve this equation, initiating $V(0)=1$, if we approximate the prefactor in the bracket on the right-hand side of Eq.~\ref{eqn_laplaceV} by  $[\frac{(3-\alpha)\alpha}{1+\alpha}]$. This approximation is valid over large range of $s$ and $\alpha$, yielding the closed expression
\begin{equation}
\tilde{V}=\frac{(1-\alpha)\gamma s^{\frac{1+\alpha}{3-\alpha}}+1}{(1+\alpha)\gamma s^{\frac{1+\alpha}{3-\alpha}} +1},
\end{equation}
where $\gamma$ is numerically determined for specific values of $\alpha$. The inverse Laplace transform is analytically tractable and results in an elaborate expression that is dominated by an exponential decay,
\begin{equation}
P(Q)\propto e^{-\frac{(3-\alpha)Q}{\gamma(\alpha+1)^2}}.
\end{equation}
Thus, we show that a combination of two singular and a continuous distributions in flow fractions gives rise to an exponential decay generalizing results previously obtained in a similar description of force distributions in random bead packs \cite{Coppersmith:96}.

At last we question how well the pore size distribution could predict flow distribution. We simulate flow through a porous medium generated by discs on a square lattice at medium positional disorder $\sigma=0.1$ with optional disorder in disc size of up to 20\% disc diameter \cite{supplement}. We find  almost identical pore size distributions for without or with disorder in disc size yet entirely different flow distributions, Gaussians versus close to exponential. A fact that is reflected very well in the corresponding characteristic distribution of fractions of propagated flow. Note that two Gaussians in the flow rate distribution translate to a single Gaussian in the flow velocity distribution. This result explains previously puzzling observations where very similar pore size distributions correspond to a Gaussian or an exponential flow distribution \cite{Maier:98}. Close inspection of the positional correlations of spherical obstacles in the previous work \cite{Maier:98}, reveal a high degree of ordering in the sample with a Gaussian flow velocity distribution in agreement with our findings.

To summarize, mapping a porous medium to a network of tubes of varying diameter, we show that a mean field description of the propagation of flows at node points within this tube network successfully predicts the flow distribution from the distribution of propagated flow fractions. The flow fractions are determined by local correlations of the pore sizes.  This demonstrates that instead of the overall pore size distribution, the local correlation of pore sizes on the scale of the pore network sets the flow distribution. Changing the local correlation of pore sizes by gradually changing the disorder in the packing of discs transitions the flow distribution from double-$\delta$-function-like at low disorder via broadened Gaussians to an exponential distribution already at a fairly moderate disorder. While flow distribution properties such as mean width or decay rate depend on network and obstacle geometry and porosity \cite{Datta:13,Matyka:2016}, distribution form and transitions only depend on the amount of disorder in the local pore correlations and are independent of geometry. Our analytical description is based on the flows within the tubes of the pore network and therefore independent of the precise dynamics of fluid flow within each tube, i.e.~two-dimensional versus three-dimensional Poiseuille flow. Given the success of a previous similar network of tubes models \cite{Bryant:1992} to describe flows through sandstone we are confident that also porous media in three dimensions should be well described as a network of tubes at low porosity. We therefore expect the general statement that local pore size correlations determine the flow distribution to carry over to three-dimensional porous media.

\begin{acknowledgments}
We thank S.~S.~Datta for initial discussions and H.~Sizaret for his contributions to the experiment. This research was funded by the Human Frontiers Science Program through Grant RGP0053/2012, the National Science Foundation (DMR-1310266), the Harvard Materials Research Science and Engineering Center (DMR-0820484) and (DMR-1420570), Total E\&P Recherche Developpement (A21116) and the Deutsche Akademie der Naturforscher Leopoldina (K.A.).  M.P.B.~acknowledges the support of the Simons Foundation.
\end{acknowledgments}
%
\newpage
\includepdf[pages={1,{}},pagecommand={},width=\textwidth]{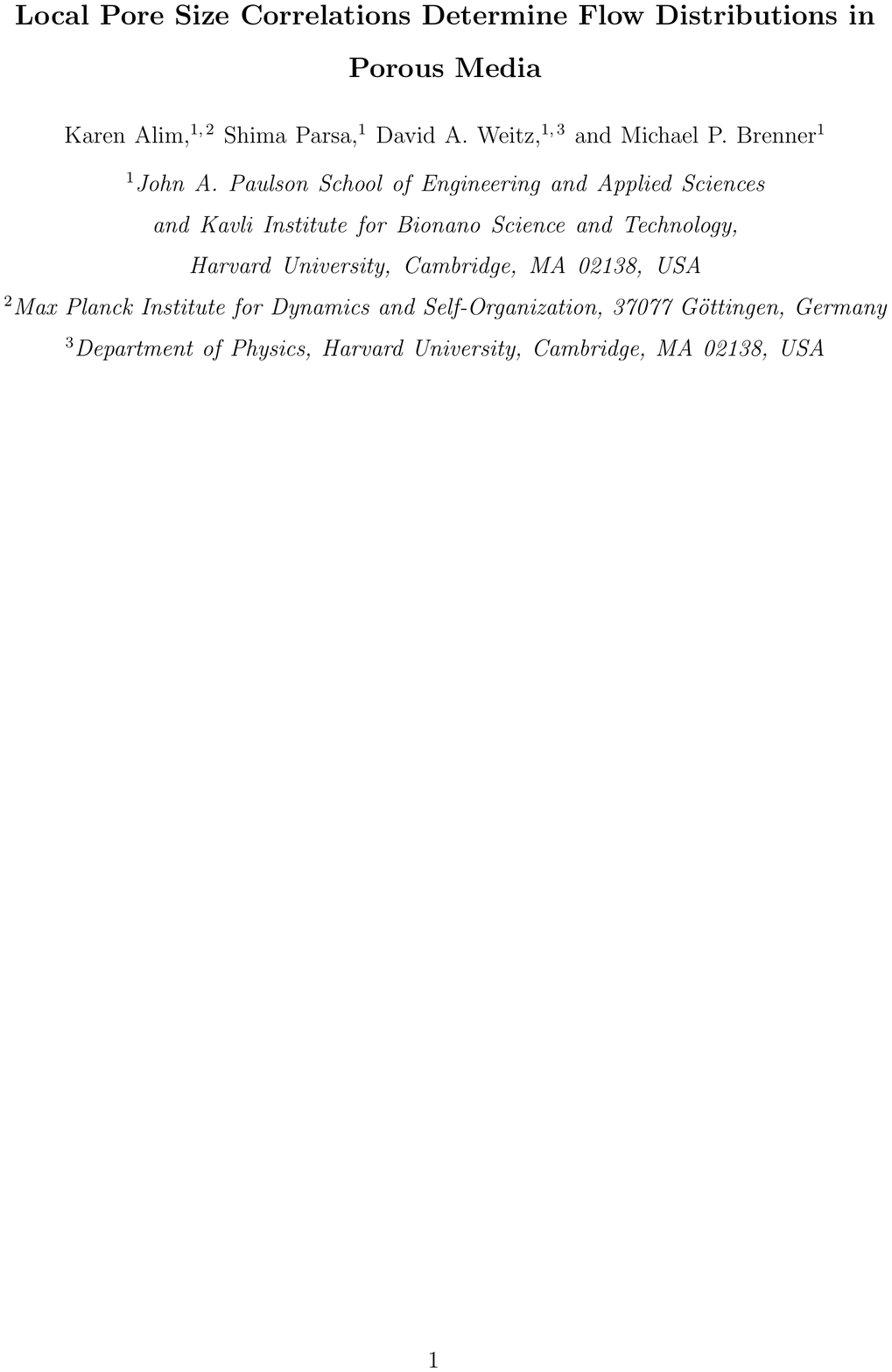}
\newpage
\includepdf[pages={2,{}},pagecommand={},width=\textwidth]{supplemental_porousmedium_arxiv.pdf}
\newpage
\includepdf[pages={3,{}},pagecommand={},width=\textwidth]{supplemental_porousmedium_arxiv.pdf}
\newpage
\includepdf[pages={4,{}},pagecommand={},width=\textwidth]{supplemental_porousmedium_arxiv.pdf}
\newpage
\includepdf[pages={5,{}},pagecommand={},width=\textwidth]{supplemental_porousmedium_arxiv.pdf}
\newpage
\includepdf[pages={6},pagecommand={},width=\textwidth]{supplemental_porousmedium_arxiv.pdf}
\end{document}